          [2001/04/25 1.1 (PWD)]
\documentclass[a4paper,twocolumn]{esapub} 
\usepackage{natbib}
\usepackage{graphicx}

\title{From BeppoSAX to INTEGRAL: PDS observations of hard X-ray sources detected in the IBIS survey}
\author{A. Malizia(1), L. Bassani(1), R. Landi(1), M. Molina(2), J.B. Stephen(1), F. Gianotti(1), \\
F. Schiavone(1), E. J. Barlow(3), A. Bazzano(4), A. J. Bird(3), F. Capitanio(4), \\ A. J. Dean(3),
M. Del Santo(4), F. Lebrun(5),  M. Renaud(5),  S. E. Shaw(3), R. Terrier(5), \\ P. Ubertini(4), 
R. Walter(6)}
\affil{(1) IASF-BO CNR/INAF Bologna, Italy }
\affil{(2) Universit\'a degli Studi di Bologna, Bologna, Italy }
\affil{(3) Phys. Dept, University of Southampton, Southampton, UK} 
\affil{(4) IASF-Ro CNR/INAF Roma, Italy}
\affil{(5) CEA Saclay/DSM/DAPNIA/Sap, Gif Sur Yvette, France}
\affil{(6) INTEGRAL Science Data Center, Geneva, Switzerland}

\begin{document}

\keywords{hard X-ray sources; compact objects}

\maketitle

\begin{abstract}
The first IBIS galactic plane survey has provided a list of high energy emitting objects
above 20 keV; these sources have been detected mostly in the
crowded region of the Galactic Centre and partly along the Galactic 
Plane. In order to validate the detection procedure, to help in the
identification process and to study the nature of these IBIS sources,
this list has been cross correlated with the data archive of the PDS
instrument on BeppoSAX, which operated in a similar energy band and with a similar
sensitivity. We discover a number of associations whose detailed
analysis will be particularly useful for the survey work. Also, 
thanks to the imaging capability of IBIS/ISGRI, objects which could not be
studied by the PDS due to contamination from nearby sources can now be
associated with a definite source or sources.
\end{abstract}

\section{Introduction}
The IBIS/ISGRI Survey (Bird et al. 2004) contains 123 high energy
emitting objects detected with the unprecedented sensitivity of
$\sim$1 mCrab in the energy range 20-100 keV discovered by mosaicing
all core program observations performed in the first year of the
mission; this first catalogue contains 23 high mass and 53 low mass
binary systems, 5 AGN, a few SNR/X-ray pulsar systems, a few isolated
pulsars and a handful of other objects. Around 30 remain at the
moment unidentified and are the main targets of follow up
observations. Observations of these and other sources in the IBIS
survey at X-ray wavelengths are useful in order to assess their nature and
overall characteristics; in this sense the BeppoSAX/PDS archive is a
powerful tool as it can provide information on any spectral and/or
flux variation provided that the PDS observation was not contaminated
by nearby sources. To this end, we have started a program to analyse
all PDS observations which contain in the field of view a source detected 
by IBIS: the systematic search of the archive has provided a set of 68
objects which were targets of BeppoSAX observations and so have both
MECS/PDS (2-100 keV band) data. 
In this case it is possible to
reconsider the PDS data in the light of the IBIS images in order to
exclude or evaluate any contamination present.  The PDS field of view
is 1.3$^{\circ}$ (FWHM), hexagonal in shape and with no imaging
capability (Frontera et al. 1997). 
The response matrix of this instrument is triangular in
both directions, with a flat top of 3' and a reduction in sensitivity
of a factor of 2 at 38' from the centre up to zero response at 78'
distance . 
The MECS has instead a field of view
of 30' radius and so covers only about 10\% of the PDS view. Therefore
it is also possible that a source is not seen by the MECS but
serendipitously observed by the PDS.  In fact, from our search, 21
IBIS survey sources were measured by the PDS in this mode.  For the
remaining sources there is no BeppoSAX pointing.  Herein we
present a few interesting cases where IBIS/PDS data can be used
together in order to define the high energy characteristics of some of
these objects which were serendipitously detected by Beppo/SAX.

\section{Data Analysis}
The dataset used in this work consists of all Core Program
observations (GPS and GCDE) performed from revolution 20 to revolution
145.  ISGRI images were generated for each selected pointing in the
sky in the 20-100 keV band by combining in mosaics all
available science windows (scw) with the ISDC offline scientific analysis software (OSA)
version 3 (Goldwurm et al. 2003).  Source ghosts in the images have been
cleaned using an appropriate catalogue available at the ISDC. More
details on the data analysis can be found in Bird et al. (2004).\\
Figure 1 shows the first field studied in this work: it represents a
mosaic in the 20-100 keV band of the Scutum region consisting of 400 science windows 
(fixed pointings that last about 37 minutes each).
In this region we find three PDS pointings: the two SNRs,
Kes75 and SNR 021.5-00.9 (case A and B) and the serendipitous PDS
detection of a new unidentified X-$\gamma$ ray source AX J1838-0655
(case C).  The two circles in figure 1 represent the PDS field of view
pointed at the two SNRs, while the case of the new source is discussed
further in a separate section.  Finally, in the last section, we
discuss the case of another new ISGRI source IGR J16207-5129 and
present its PDS spectrum. It is important to note that none of the
BeppoSAX data presented in this paper have been reported in the
literature yet.\\
In the following, ISGRI fluxes are based on the count rates quoted in the survey
(Bird et al. 2004) and scaled with the Crab results.

\begin{small}
\begin{figure}
\centering
\includegraphics[width=0.8\linewidth]{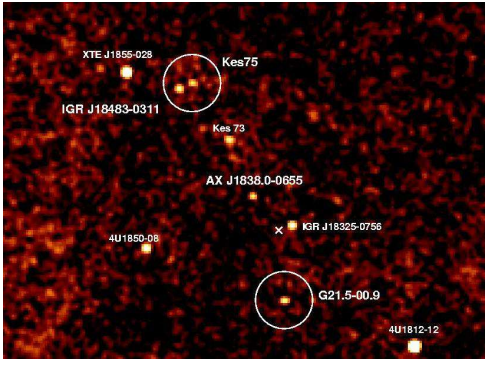}
\caption{IBIS/ISGRI (20-100 keV) mosaic of the Scutum region obtained by summing
400 science windows ($\sim$880 ksec). The circles represent the BeppoSAX/PDS field of view. 
The X-point is one of the two offset pointings of G021.5-00.9 observation.\label{fig:single}}
\end{figure}
\end{small}

\begin{small}
\begin{figure}
\centering
\includegraphics[width=0.8\linewidth,angle=-90]{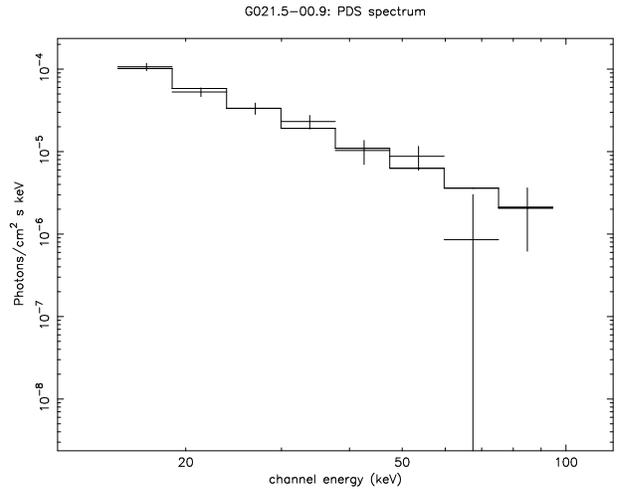}
\caption{PDS 20-100 keV corrected spectrum of the IBIS Survey source G21.5-09\label{fig:single}}
\end{figure}
\end{small}

\subsection{Case A: SNR 021.5-00.9}
BeppoSAX pointed at the SNR  021.5-00.9 three times and high energy emission
(20-100 keV) has been detected by the PDS instrument.
This looks as the simplest case of figure 1 where the only high 
energy emitter in the PDS field of view is SNR  021.5-00.9. 
Since the PDS detection looks uncontaminated we have analysed the data:
a fit with a simple power law ($\Gamma$=2.4)
gives a 20-100 keV average flux of $\sim$ 1.2 $\times$ 10$^{-11}$ erg
cm$^{-2}$ s$^{-1}$.
This  flux is lower than that seen by ISGRI (Bird et al. 2004) indicating a
possible variability of the source or a contamination in the PDS offset
pointings.
Since variability in unlikely in  G021.5-00.9, we have checked if the offset 
field measurements provide indication for the presence of a high energy emitting source.
Indeed a 9$\sigma$ source was discovered in one of the two offset pointings:
cross-checking this PDS offset pointing with
the ISGRI image we found that the contaminating source was IGR J18325-0756
(see X point in figure 1).
The corrected spectrum of  SNR 021.5-00.9 is shown in figure 2.
A fit to this spectrum provides a $\Gamma$ of about 2.6 and a 20-100 keV flux
of 4-5 $\times$ 10$^{-11}$ erg cm$^{-2}$ s$^{-1}$ more in line with the ISGRI 
detection ($\sim$ 6  $\times$ 10$^{-11}$ erg cm$^{-2}$ s$^{-1}$ ).
Furthermore, the PDS spectrum of IGR J18325-075 (see figure 3), is well fitted with
a power law with $\Gamma$$\sim$2.5 and a 20-100 keV flux of 7 $\times$ 10$^{-11}$ erg cm$^{-2}$ s$^{-1}$
again consistent with the ISGRI detection ($\sim$ 8 $\times$ 10$^{-11}$ erg cm$^{-2}$ s$^{-1}$).

\begin{small}
\begin{figure}
\centering
\includegraphics[width=0.8\linewidth,angle=-90]{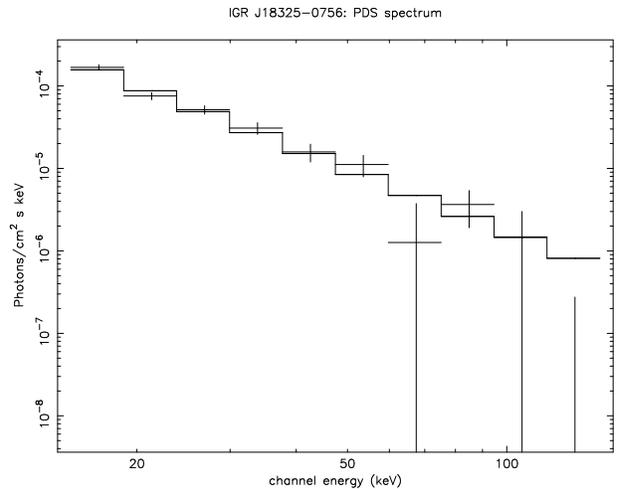}
\caption{PDS 20-100 keV of IGR J18325-0756.\label{fig:single}}
\end{figure}
\end{small}

\subsection{Case B: Kes 75 (SNR 029.7-00.2)}
As clearly shown in figure 1, in the PDS field of view of the BeppoSAX
pointing of Kes 75 (also AXJ1846.4-0258, a SNR with an X-ray pulsar)
there is another high energy emitter which was detected for the first
time by ISGRI (Bird et al. 2004) and named IGR J18483-0311. This new
ISGRI source is 31' away from Kes 75, and so it is not in the
MECS field of view but is likely to have been serendipitously detected
by the PDS, unless it is a transient object not active at the time of
the BeppoSAX measurement.  
Analysis of the BeppoSAX of Kes 75 indicates contamination 
by this nearby source and in fact the MECS/PDS
cross-calibration constant is 3.5$^{+0.5}_{-0.3}$ i.e.  outside the
nominal range of 0.75-0.95 generally observed.  In figure 4 the PDS
spectrum from the pointing containing both Kes 75 and IGR J18482-0311 is
reported. The solid line is the extrapolation of the MECS spectrum of
Kes 75 and provides 20-100 keV flux of 3 $\times$ 10$^{-11}$ erg cm$^{-2}$ s$^{-1}$.
The difference between this and the data points is reported in figure 5
which should be taken with caution but can be use as a reference spectrum for this ISGRI source.
The estimated flux is $\sim$6 $\times$ 10$^{-11}$ erg
cm$^{-2}$ s$^{-1}$ consistent with the ISGRI detection ($\sim$ 6 $\times$ 10$^{-11}$ erg cm$^{-2}$ s$^{-1}$).

\begin{small}
\begin{figure}
\centering
\includegraphics[width=0.8\linewidth,angle=-90]{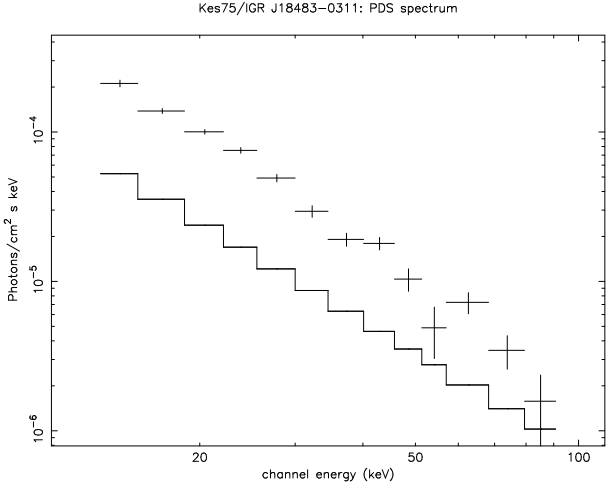}
\caption{PDS 20-100 keV spectrum of the pointing containing the IBIS Survey source Kes 75 (SNR)
which is clearly contaminated by a nearby ISGRI source IGR J18483-0311.
The solid line is the extrapolation of the MECS spectrum of Kes 75.\label{fig:single}}
\end{figure}
\end{small}

\begin{small}
\begin{figure}
\centering
\includegraphics[width=1.1\linewidth]{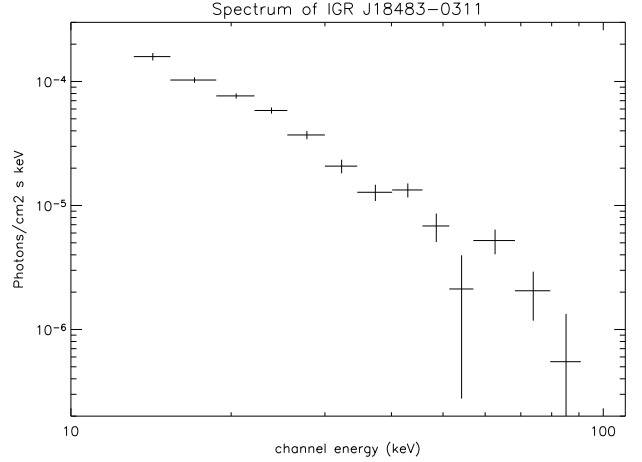}
\caption{PDS 20-100 keV spectrum of the new ISGRI source IGR J18483-0311.\label{fig:single}}
\end{figure}
\end{small}

\begin{small}
\begin{figure}
\centering
\includegraphics[width=0.8\linewidth]{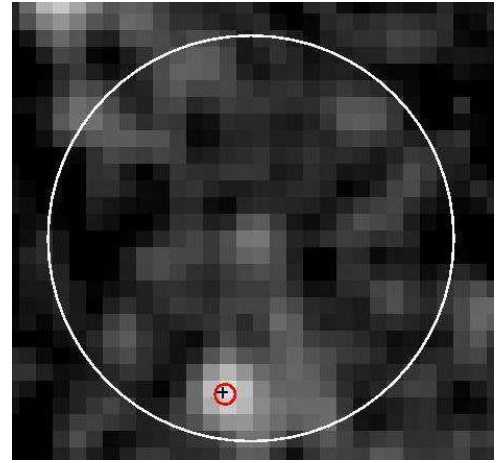}
\caption{A zoom of the central region of figure 1 on the X-ray source AX J1838.0-0655.
Einstein (cross point) and ASCA (small circle) are superimposed, the large circle represent 
the PDS field of view. \label{fig:single}}
\end{figure}
\end{small}

\begin{small}
\begin{figure}
\centering
\includegraphics[width=0.8\linewidth,angle=-90]{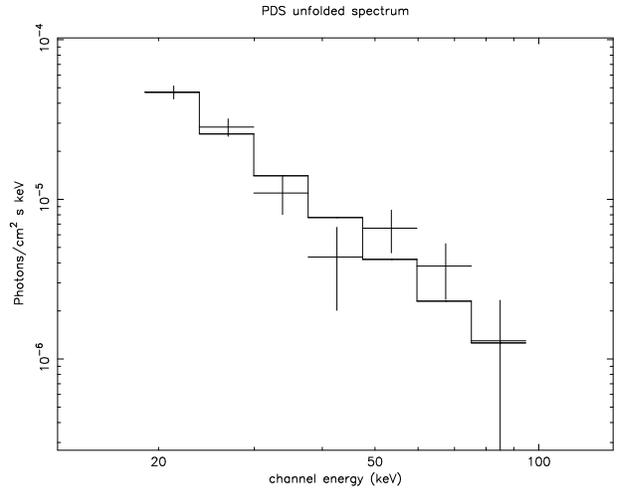}
\caption{PDS spectrum of AX J1838.0-0655.\label{fig:single}}
\end{figure}
\end{small}

\subsection{Case C: AX J1838.0-0655}
In figure 6 a close-up of the area around the IBIS survey source AX
J1838.0-0655 (see also figure 1) is shown.  AX J1838.0-0655 was first
detected by the IPC instrument on the Einstein satellite
(1E1835.3-0658, Hertz \& Grindlay 1988) with an associated 90\% error
radius of 49''. The source was also detected by ASCA in the 0.7-10 keV
range during the galactic plane survey (Sugizaki et al. 2001 and Bamba
et al. 2003) with an associated uncertainty of 1' radius. Finally,
AX J1838.0-0655 was serendipitously detected twice by the PDS instrument
during the BeppoSAX pointings devoted to the study of the GEV source
1837-0610. In figure 6 the cross represents the Einsten
position, the small circle the ASCA error box while the large circle
is the PDS field of view. It is clear from the IBIS image that AX
J1838-0655 is the only hard X-ray emitter in the PDS field of view. It
is detected by ISGRI at 13$\sigma$ in the 20-100 keV range in the
cumulative exposure with an average flux of about 2-3 mCrab.  The
source, still unidentified, is a bright (1.1 $\times$ 10$^{-11}$ erg
cm$^{-2}$ s$^{-1}$, flat ($\Gamma$=0.5) and variable ASCA source
(Sugizaki et al. 2001). The PDS data show a steep spectrum
($\Gamma$=2.5, see figure 7) and a 20-100 keV flux of around 2 mCrab
in agreement with the ISGRI flux.  A detailed discussion on the
possible nature of AX J1838.0-0655 is reported in Malizia et al. (2004).

\section{IGR J16207-5129}
In figure 8 the ISGRI image (20-100 keV) of the field containing
another new IBIS Survey source IGR J16207-5129
(Atel \#229, Walter et al. 2004) is shown. The image is a zoom of a mosaic of the Norma
region obtained by summing 470 science windows.  Again the circle
represents the PDS field of view.  In this last case, BeppoSAX was
targeted on the SNR 1E161348-5055 which is located at the centre of the
circle and which obviously has not been detected by ISGRI.  BeppoSAX
pointed at 1E161348-5055 three times in 1998/99 and a source was
detected in the first and third pointings only, demonstrating
variability over a period of 6 months.  The high energy emission
definitely does not come from the SNR since the MECS/PDS
cross-calibration factor is higher than 100.  In both the first and
the last observations, the 20-100 keV data are well fitted with a simple
power law of photon index $\Gamma$$\simeq$2.5 with a flux of $\sim$ 6
$\times$ 10$^{-11}$ erg cm$^{-2}$ s$^{-1}$, much in agreement with the
ISGRI data ($\leq$ 7 $\times$ 10$^{-11}$ erg cm$^{-2}$ s$^{-1}$).  In figure 9 the PDS 20-100 keV
spectrum of the first pointing is shown.  The ISGRI data
demonstrates that the source of the high energy emission is not the
BeppoSAX target, but is the new ISGRI source IGR J16207-5129.

\begin{small}
\begin{figure}
\centering
\includegraphics[width=0.8\linewidth]{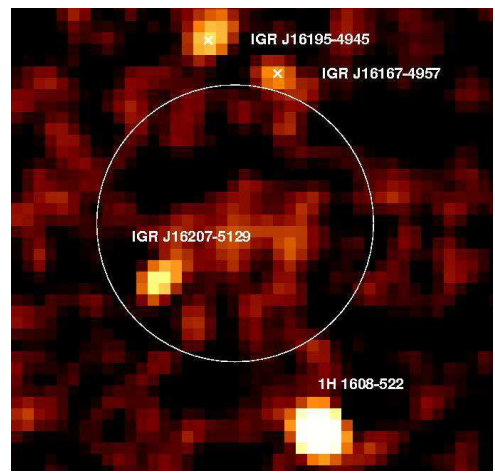}
\caption{A close-up of the area in the Norma region containing the
new ISGRI source IGR J16207-5129. The mosaic is obtained by summing 470
science windows in the 20-100 keV energy band. \label{fig:single}}
\end{figure}
\end{small}

\begin{small}
\begin{figure}[t]
\centering
\includegraphics[width=0.8\linewidth,angle=-90]{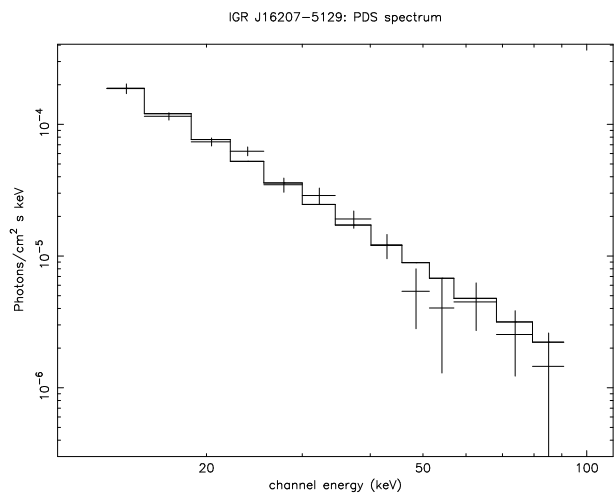}
\caption{PDS spectrum of IGR J16207-5129.\label{fig:single}}
\end{figure}
\end{small}

\section{Conclusions}
In this paper we have demonstrated that the combination of BeppoSAX/PDS and ISGRI data can produce
a clearer understanding of the nature of high energy emitters. Work is
continuing on this correlation analysis and will improve as the IBIS
survey dataset increases.

\section*{Acknowledgments}
We acknowledge financial support by ASI (Italian Space Agency) via contract
I/R/041/02.


\end{document}